\documentclass[fleqn,usenatbib]{mnras}

\usepackage{newtxtext,newtxmath}

\usepackage[T1]{fontenc}

\DeclareRobustCommand{\VAN}[3]{#2}
\let\VANthebibliography\thebibliography
\def\thebibliography{\DeclareRobustCommand{\VAN}[3]{##3}\VANthebibliography}

\usepackage{graphicx}	        
\usepackage{amsmath}	        
\usepackage{bm}                 
\usepackage{float}
\usepackage{caption}            
\usepackage{subcaption}
\usepackage{color}              
\usepackage[normalem]{ulem}     
\usepackage{nicefrac, xfrac}    
\usepackage{makecell}

\newcommand{\CD}[1]{\textcolor{cyan}{#1}}

\graphicspath{{Figures/}}

\title[Modelling Force-Free Magnetospheres using PINNs]{Modelling Force-Free Neutron Star Magnetospheres using Physics-Informed Neural Networks}

\author[J. F. Urb\'an et al.]{
Jorge F. Urb\'an, $^{1}$\thanks{E-mail: jorgefrancisco.urban@ua.es}
Petros Stefanou,$^{1,2}$\thanks{E-mail: petros.stefanou@uv.es}
Clara Dehman$^{3,4}$, Jos\'e A. Pons$^{1}$
\\
$^{1}$Departament de Física Aplicada, Universitat d'Alacant, Ap. Correus 99, E-03080 Alacant, Spain\\
$^{2}$Departament d'Astronomia i Astrofísica, Universitat de València, Dr. Moliner 50, E-46100, Burjassot, València, Spain\\
$^{3}$Institute of Space Sciences (ICE-CSIC), Campus UAB, Carrer de Can Magrans s/n, 08193, Barcelona, Spain\\
$^{4}$Institut d'Estudis Espacials de Catalunya (IEEC), Carrer Gran Capità 2–4, 08034 Barcelona, Spain\\
}

\date{Accepted XXX. Received YYY; in original form ZZZ}

\pubyear{2023}

\begin{document}
\label{firstpage}
\pagerange{\pageref{firstpage}--\pageref{lastpage}}
\maketitle

\begin{abstract}

Using Physics-Informed Neural Networks (PINNs) to solve a specific boundary value problem is becoming more popular as an alternative to traditional methods. However, depending on the specific problem, they could be computationally expensive and potentially less accurate. The functionality of PINNs for real-world physical problems can significantly improve if they become more flexible and adaptable. To address this, our work explores the idea of training a PINN for general boundary conditions and source terms expressed through a limited number of coefficients, introduced as additional inputs in the network. Although this process increases the dimensionality and is computationally costly, using the trained network to evaluate new general solutions is much faster. Our results indicate that PINN solutions are relatively accurate, reliable, and well-behaved. We applied this idea to the astrophysical scenario of the magnetic field evolution in the interior of a neutron star connected to a force-free magnetosphere. Solving this problem through a global simulation in the entire domain is expensive due to the elliptic solver's needs for the exterior solution. The computational cost with a PINN was more than an order of magnitude lower than the similar case solved with classical methods. These results pave the way for the future extension to 3D of this (or a similar) problem, where generalised boundary conditions are very costly to implement.

\end{abstract}




\section{Introduction}

Deep learning (DL) is a subset of techniques comprehended in Machine Learning that is fundamentally based on multi-layered Neural Networks (NNs). In recent years, DL has been widely used to perform a large variety of tasks. Examples include (among many others): computer vision (to perform image classification) \citep{Traore_Kamsu-Foguem_Tangara_2018}, face recognition \citep{Lawrence_Giles_Tsoi_Back_1997} or medical diagnosis \citep{Kugunavar_Prabhakar_2021}, speech recognition \citep{Chan_Jaitly_Le_Vinyals_2015} and robotics to emulate human-like walking and running or mobile navigation in pedestrian environments \citep{Hayat_Mall_2013}.

Physics-Informed Neural Networks (PINNs) \citep{Raissi_Perdikaris_Karniadakis_2019}, is a deep learning approach used to numerically approximate the solution of non-linear partial differential equations (PDEs). The original idea was born more than twenty years ago \citep{Lagaris_Likas_Fotiadis_1997}, but the lack of the necessary computational resources made it complicated to put it into practice. In recent years, we account with graph-based automatic differentiation, as well as different frameworks that support computations in CPUs and GPUs, such as Tensorflow or Pytorch, and a dramatic increase in computational power. These factors, combined with a blooming interest in machine learning applications in science, have given birth to this promising new field. PINNs have been used, among many other applications, in fluid dynamics \citep{Cai_Mao_Wang_Yin_Karniadakis_2021}, nuclear reactor dynamics \citep{Schiassi_DeFlorio_Ganapol_Picca_Furfaro_2022}, radiative transfer \citep{Mishra_Molinaro_2021, Chen_Jeffery_Zhong_McClenny_Braga-Neto_Wang_2022} and black-hole spectroscopy \citep{Luna_Calderon-Bustillo_Seoane-Martinez_Torres-Forne_Font_2022}.
PINNs incorporate the underlying physical laws that govern a system (the PDEs) in the loss function and then optimise the NN so that the residual of the PDE is minimal. Unlike more traditional DL approaches in other fields, PINNs do not require large amounts of data --or any data at all-- for the training of the NN. 

Compared to classical finite-differences/finite-elements methods, PINNs still fall short in terms of efficiency and precision. However, they present some advantages as flexible, multi-purpose PDE solvers. For example, the PINN formulation allows us to solve problems in arbitrary, unstructured meshes without using high-resolution, memory-consuming grids. In addition, once a PINN is trained for a general problem, the calculation of a new solution is swift and consists only of the few operations needed during the forward pass through the network. This is a potential advantage in speed compared to classical methods. 

In this work, we assess the applicability of a PINN solver in elliptic problems. In particular, we focus on the problem of modelling force-free (FF) magnetospheres of neutron stars (NS) in the non-rotating, axisymmetric limit. There is a wealth of work in the literature formulating this problem in terms of the Grad--Shafranov equation \citep{Glampedakis_Lander_Andersson_2014, Pili_Bucciantini_Del_Zanna_2015, Akgun_Miralles_Pons_Cerda_2016, Kojima_2017, Akgun_Cerda_Miralles_Pons_2017a, Akgun_Cerda_Miralles_Pons_2017b, Akgun_Cerda_Miralles_Pons_2018}, which gives us the opportunity to make detailed comparisons and draw robust conclusions on the performance and generalisability of the PINN solver. 

The paper is organized as follows: in Sec.~\ref{sec:basic_equations}, we give a brief mathematical overview of the physics of NS magnetospheres. In Sec.~\ref{sec:pinn_solver}, we describe in detail how the PINN solver is built. We present the solutions acquired by the PINN solver with error estimates in Sec.~\ref{sec:tests_and_models}. In Sec.~\ref{sec:application_to_MT}, we demonstrate PINN's capabilities through an astrophysical application. Sec.~\ref{sec:conclusions} is dedicated to discuss our main results and how to improve and generalise our approach to face more difficult problems in the near future.

\section{Modelling axisymmetric force-free magnetospheres}\label{sec:basic_equations}
In the force-free regime, and considering that the contribution of gravity, inertia, plasma pressure and rotation in the dynamics of a NS magnetosphere is negligible compared to the magnetic force, the force-balance equation reduces to the simple form
\begin{equation} \label{eq:ff_equation}
    (\bm{\nabla} \times \bm{B}) \times \bm{B} = 0, 
\end{equation}
where $\bm{B}$ denotes the magnetic field.
This regime is better suited for  magnetar magnetospheres \citep{Thompson_Duncan_1995, Thompson_Duncan_1996}, because magnetars are slow rotators and the absence of any rotationally induced electric fields is a very good approximation. 

In axisymmetry, we can express the magnetic field in terms of a poloidal and a toroidal stream function ${\cal P}$ and ${\cal T}$. We will follow the notation and formalism as in \citet{Akgun_Miralles_Pons_Cerda_2016}.
We refer the interested reader to that work for a more detailed description.
In spherical coordinates $(r,\theta, \phi)$ and in terms of the stream functions the magnetic field reads:
\begin{equation} \label{eq:poloidal_toroidal_decomposiosion}
    \bm{B} = \bm{\nabla} {\cal P} \times \bm{\nabla} \phi + {\cal T} \bm{\nabla} \phi.
\end{equation}
where $\bm{\nabla} \phi = \frac{\bm{e}_\phi}{r \sin\theta}$ with $\bm{e}_\phi$ being the azimuthal unit vector. Substituting Eq.~\eqref{eq:poloidal_toroidal_decomposiosion} into~\eqref{eq:ff_equation}, the $\phi-$component of the equation gives
\begin{equation}
    \bm{\nabla} {\cal P} \times \bm{\nabla} {\cal T} = 0, 
\end{equation}
which simply states that, ${\cal T} = {\cal T}({\cal P})$ must be a function of ${\cal P}$ 
(or vice-versa).
The remaining components, give us the so-called Grad--Shafranov (GS) equation
\begin{equation} \label{eq:grad_shafranov}
    \triangle_{\text{GS}} {\cal P} + G({\cal P})=0.
\end{equation}
Here $G(\mathcal{P}) = {\cal T}({\cal P}) \frac{d {\cal T}}{d{\cal P}}$ is the source term accounting for the presence of currents in the magnetosphere and $\triangle_{\text{GS}}$ is the GS operator
\begin{equation} 
    \triangle_{\text{GS}} \equiv r^2 \sin^2\theta ~\bm{\nabla} \cdot 
    \left( \frac{1}{r^2 \sin^2\theta} \bm{\nabla}\right) ~.
\end{equation}

For convenience, we will use compactified spherical coordinates (see \cite{Stefanou23}) $(q, \mu, \phi)$, where $q = \frac{1}{r}$ and $\mu = \cos{\theta}$ instead of the usual $(r, \theta, \phi)$. In this set of coordinates the GS operator reads:
\begin{equation} \label{eq:grad_shafranov_operator}
    \triangle_{\text{GS}} \equiv q^2 \partial_q \left(q^2 \partial_q \right) + \left(1-\mu^2 \right)q^2 \partial_{\mu \mu}. 
\end{equation}

To solve Eq.~\eqref{eq:grad_shafranov}, we must also provide boundary conditions (BCs) and the functional form of the source term, that is, ${\cal T}({\cal P})$. 
The particular functional form is arbitrary and different choices are possible. 
In \citet{Akgun_Miralles_Pons_Cerda_2016}, they used
\begin{equation}
    {\cal T}({\cal P}) = s \left({\cal P} - {\cal P}_c \right)^\sigma 
    \Theta\left({\cal P}-{\cal P}_c\right),  
\end{equation}
where $\Theta$ is the Heaviside function, and $s$, ${\cal P}_c$, and $\sigma$ are parameters that control the relative strength of the toroidal and poloidal components, the region where the toroidal field is non-zero and the non-linearity of the model.
However, this has the limitation that it assumes ${\cal P}>0$. A possible generalisation
overcoming this constraint is 
\begin{equation}\label{eq:T_of_P}
    {\cal T}({\cal P}) = s \left(\left|{\cal P}\right| - {\cal P}_c \right)^\sigma 
    \Theta\left(\left|{\cal P}\right|-{\cal P}_c\right),  
\end{equation}
which allows for currents even for negative values of ${\cal P}$.
We have explored different options but, for simplicity and the purpose of this paper, we will use a quadratic function for the astrophysical application in Sec.~\ref{sec:FF evolution}, defined as follows:
\begin{equation}\label{eq:T_of_P_quadratic}
    \mathcal{T}(\mathcal{P}) = s_1 \mathcal{P} + s_2 \mathcal{P}^2.
\end{equation}
We impose BCs for ${\cal P}$ at the surface of the star $(q=1)$, at radial infinity $(q=0)$ and at the axis $(\mu = \pm 1)$. Regularity and symmetry of the problem lead to 
$${\cal P}(q, \mu = \pm 1) = {\cal P} (q=0, \mu) = 0~.$$ 
In particular, one advantage of compactifying the radial coordinate (going from $r$ to $q)$ is to make it easier to impose BCs at radial infinity: rather than imposing a specific decay rate at large $r$, we can impose Dirichlet BCs at just one point ($q=0$). This reduces unwanted numerical noise from the external boundary.

At the surface, we must provide the function ${\cal P}(\mu)$. Our implementation of BCs in a NN must be as general as possible but keep the number of parameters reasonably low. A reasonable and practical choice is to use some decomposition of the arbitrary function in terms of orthonormal polynomials. Considering the symmetry of our problem and that we are working with functions describing magnetic fields, the natural choice is to express ${\cal P}(q=1)$ in terms of coefficients of a Legendre polynomial expansion. We 
use the following decomposition:
\begin{equation}\label{eq:Psurface}
    {\cal P} (q=1, \mu) = \left(1-\mu^2\right) \sum_{l=1}^{l_{\text{max}}} 
    \frac{b_l}{l} P_l'(\mu),
\end{equation}
where $l$ is the order of the multipole ($l=1$ corresponds to a dipole), $P_l$ are the Legendre polynomials (not to be confused with ${\cal P}$, the poloidal flux function) and the prime denotes differentiation with respect to $\mu$. Thus, the boundary condition at the surface is completely determined by prescribing the $b_l$ coefficients \footnotemark.

\footnotetext{The $1/l$ normalisation factor and the coefficients $b_l$ in the expansion have been chosen to match the $b_l$ coefficients used in \cite{Dehman_Vigano_Pons_Rea_2023}.}

\section{Methodology}\label{sec:pinn_solver}

\subsection{Neural Networks}

NNs are universal approximators of mathematical functions \citep{Hornik_Stinchcombe_White_1989}. 
They are the result of compositions of simple but non-linear transformations at different layers. 
The way that these layers are interconnected indicates the NN \emph{architecture}. 
There are many architectures available in the literature, such as Fully-Connected Neural Networks (FCNNs), Recurrent Neural Networks (RNNs), Convolutional Neural Networks (CNNs) and more. 
Each one is particularly suitable for specific tasks (see \citealt{Alzubaidi2021ReviewOD} for a detailed review).
In this paper, we have adopted two different types of architecture: FCNNs and Residual Neural Networks (ResNets). We briefly describe these architectures below.

\subsubsection{Fully-Connected Neural Networks}

\begin{figure}
    \centering
    \includegraphics[width=\columnwidth]{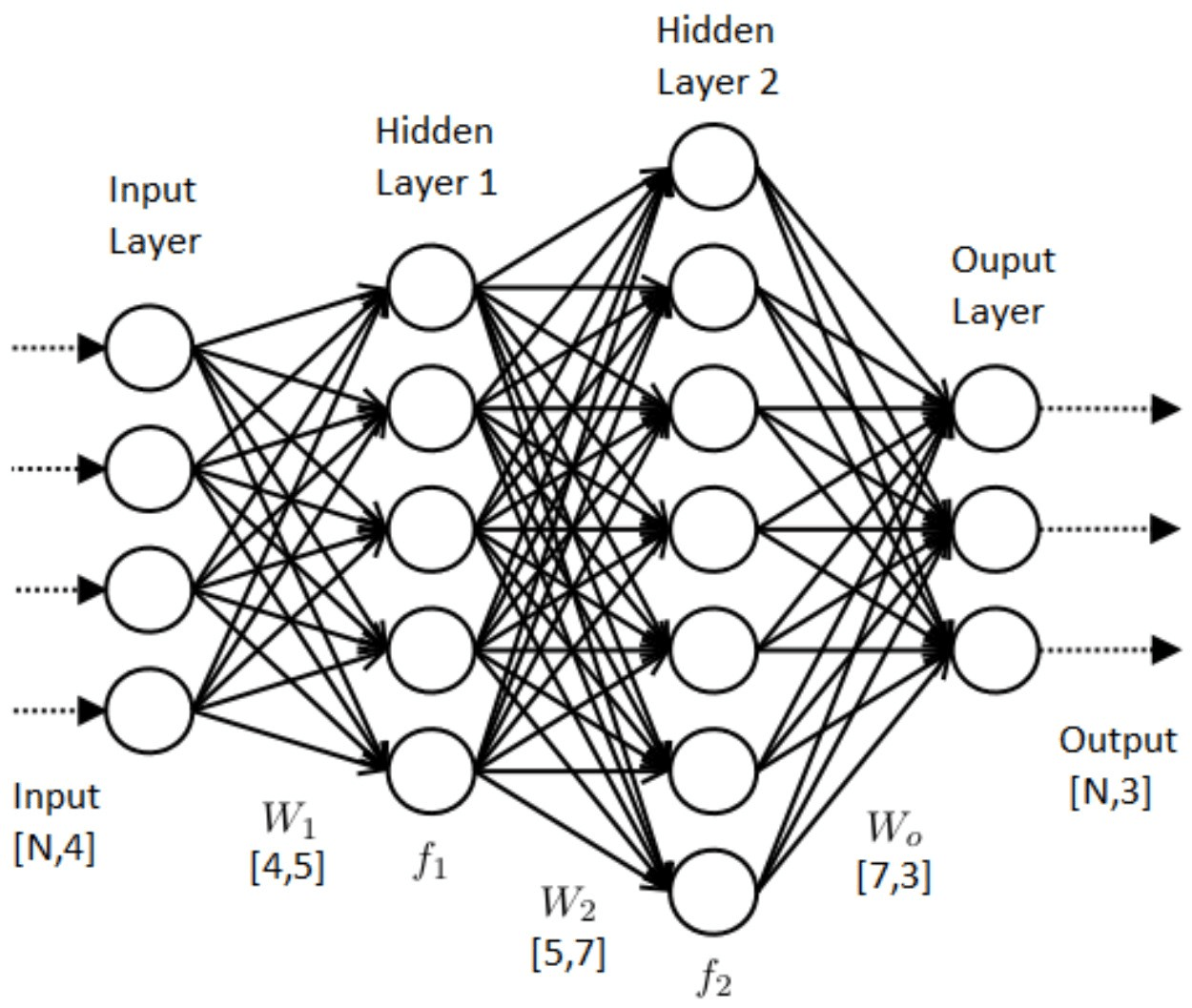}
    \caption{A schematic representation of a deep fully-connected neural network.}
    \label{fig:fcnn_sketch}
\end{figure}

In a FCNN, each layer contains a number of \emph{units} (also called neurons) that transform the inputs received from the previous layer and then pass the result to the next layer. 
This transformation is done by two basic steps. 
A linear combination of the inputs received and an evaluation through a non-linear \emph{activation function}. Mathematically, this process can be expressed for each layer as follows:
\begin{equation} \label{eq:fcnn_forward_pass}
    \bm{a}^j = g (\bm{\mathcal{W}}^j \bm{a}^{j-1} + \bm{b}^j)
\end{equation}
where $\bm{a}$ is a vector containing the values of the neurons \textbf{at} \CD{in} layer $j$. 
$\bm{\mathcal{W}}$ and $\bm{b}$ denote the \emph{weight} matrix and the \emph{bias} vector of the layer, and $g$ is a non-linear function (the activation function). 
The weights and biases of all the layers constitute the set of trainable parameters of a FCNN, i.e., the parameters that are optimised to obtain the best approximation to the true solution for our problem. 
Fig.~\ref{fig:fcnn_sketch} shows a schematic representation of a FCNN.

\subsubsection{Residual Neural Networks}

As the complexity of the problem grows, an increased number of layers is required in order to capture all the desired features of the solution \citep{Montúfar_Pascanu_Cho_Bengio_2014}. However, training deeper NNs is not a trivial task. 
Indeed, it has been demonstrated that increasing the depth leads to \emph{degradation} of the network's accuracy \citep{Srivastava_Greff_Schmidhuber_2015}, i.e., the network performs worse.

In order to overcome this problem, \citet{He_Zhang_Ren_Sun_2015} introduced the concept of ResNets. 
The idea behind this architecture is based on the phenomenological fact that it is easier to optimise a layer to learn the residual $\mathcal{F}(\bm{x}) = \mathcal{H}(\bm{x}) - \bm{x}$ of a function $\mathcal{H} (\bm{x})$ with respect to the identity function, than to learn the function $\mathcal{H}(\bm{x})$ itself. 
This is implemented by replacing a usual layer with a \emph{residual block}.
A residual block is formed if, before the evaluation of the activation function at a certain layer $j$, we add the output of the $K_{th}$ previous layer (a \emph{skip connection}).
By adding skip connections, the output of a ResNet layer can be expressed as
\begin{equation}\label{resnet_forward_pass}
    \bm{a}^j = g\left(\bm{\mathcal{W}}^{j}\bm{a}^{j-1} + \bm{b}^j + \bm{a}^{j-K}\right).
\end{equation}
 
\subsection{Physics-Informed Neural Networks}

The standard way of training a NN accounts with data that consist of values of the true solution in a given discrete set of points in the input domain.
However, this approach demands a large number of training examples to build a reliable relationship between the inputs and the outputs. 
In particular, for astrophysical systems, this method would rely on data obtained through a large set of observations. 
Training the network means to optimise its parameters so that the difference between the NN prediction $u(\bm{x})$ and the "exact" solution $\tilde{u}(\bm{x})$ is minimised. 
This is done through the use of a \emph{loss function}, which is usually some suitable norm of the quantity $|u(\bm{x})-\tilde{u}(\bm{x})|$.

The key novelty in PINNs is the incorporation of information
about the physical laws into the training process. 
This can be accomplished by minimizing the residual of the PDEs that govern the system, instead of the difference between prediction and real data/exact solution. Using data in the loss function is optional and sometimes may facilitate the optimization, but in practise is not necessary. 
Consider a general PDE of the form
\begin{align}
    \mathcal{L}u(\bm{x}) - G(\bm{x}, u(\bm{x})) &= 0, \label{eq:general_pde}\\
    u |_{\partial \mathcal{D}} &= f_b (\bm{x}), \label{eq:general_pde_boundary}
\end{align}
where $\mathcal{L}$ is a general non-linear differential operator, $G$ is a source term and $\bm{x}$ is a vector of coordinates in some domain $\mathcal{D}$. The PDE is subject to some BCs $f_b$ at the boundary $\partial \mathcal{D}$ of the domain. 
If $u$ is an approximation to an exact solution given by a PINN, then Eq.~\eqref{eq:general_pde} will have a residual, that is, the right-hand side will not be exactly zero.
The smaller this residual is, the closer $u$ be to an exact solution. Thus, the loss function is precisely a suitable norm of the residual $|\mathcal{L}u(\bm{x}) - G(\bm{x}, u(\bm{x}))|$ of the PDE.

The function $u$ should also satisfy the BCs~\eqref{eq:general_pde_boundary}. There are different approaches to implement them. 
The most commonly used is to add a term in the loss function consisting of a norm of the residual of Eq.~\eqref{eq:general_pde_boundary}, so that the residuals of both Eq.~\eqref{eq:general_pde} and~\eqref{eq:general_pde_boundary} are minimised simultaneously. 
We believe that this is not the optimal way to impose BCs, because the relative contribution of the two terms can differ significantly, thus hindering the training process. 
Ideas to surpass this problem include adjusting hyperparameters (such as the number of boundary points considered and the relative weight of the two terms) or calculating the \emph{neural tangent kernel} of the network \citep{Wang_Yu_Perdikaris_2022}, among others.
We find that, overall, the increased complexity and the need for fine tuning additional hyperparameters in this method is an inconvenience.
Instead we opt for an approach inspired by \citet{Lagaris_Likas_Fotiadis_1997} (see also a similar one, based on distance functions in \citet{Sukumar_Srivastava_2022}).
In that approach, the approximate solution $u$ is formulated as follows:
\begin{equation}\label{eq:parametrization}
    u (\bm{x}) = f_b (\bm{x}) + h_b (\bm{x}) \mathcal{N} (\bm{x}),
\end{equation}
where $f_b$ is a smooth and (at least) twice differentiable function that satisfies the BCs (see Eq.~\eqref{eq:general_pde_boundary}), $h_b$ is a smooth and twice differentiable function that defines the boundary ($h_b=0$ at $\partial \mathcal{D}$), and $\mathcal{N}$ is the output of the network.
By parametrizing the approximate solution in such a way, we ensure that the BCs are satisfied always by construction. 
The output of the network is not directly the approximate solution $u$ to our problem, but some function $\mathcal{N}$ that, when inserted in Eq.~\eqref{eq:parametrization}, gives us an approximation that satisfies exactly the BCs.

In this work, we attempt to generalise the PINN approach to build a PDE solver valid for different and varied BCs (and possibly, source terms). 
We want our network to learn how to approximate any particular solution for a given operator $\mathcal{L}$. 
This means that the information about the BCs should be part of the \emph{input} of the network (along with the coordinates) and not hardcoded in the loss function or in the parametrization~\eqref{eq:parametrization}. 
During training, the network needs to process a large number of points $\bm{x}$ and a large number of $f_b$ functions so that it can generalise and provide solutions of~\eqref{eq:general_pde} for any point in the domain and any BC.
Of course, $f_b$ could, in principle, be an arbitrary continuous function. For this reason, it should be intelligently encoded into the network's input to keep the number of parameters small and manageable.

In the following section, we present tests for our PINN solver for the Grad--Shafranov equation as described in Sec.~\ref{sec:basic_equations}.

\section{Test and models}\label{sec:tests_and_models}

\subsection{Current-free Grad--Shafranov equation}\label{subsec:current-free}

As a first test, we consider the Grad--Shafranov equation 
(Eq.~\eqref{eq:grad_shafranov}) 
without current, that is $G(\mathcal{P})=0$. 
We set up the various elements of the PINN solver as follows:
The function $h_b$ describing the boundary is given by
\begin{equation}\label{eq:h_b}
    h_b (q, \mu) = q (1 - q) (1 - \mu^2).
\end{equation}
The function $f_b$ that satisfies the BCs, where $h_b=0$, is given by
\begin{equation}\label{eq:f_b}
    f_b (q, \mu) =  q^n \left(1-\mu^2\right)\sum_{l=1}^{l_{\text{max}}} \frac{b_l}{l} P_l'(\mu). 
\end{equation}
Notice that Eq.~\eqref{eq:f_b} differs from Eq.~\eqref{eq:Psurface} by a factor $q^n$ with $n>0$. 
We include this factor to enforce BCs both at the surface ($q=1$) and infinity ($q=0$). 
If $n=1$, this parametrization is the same as that used in \citealt{Lagaris_Likas_Fotiadis_1997} considering essential BCs in a rectangle.  
However, we prefer to leave $n$ as a free parameter that is used to give more or less weight to the solution close to the star or away from the surface. 
We performed a detailed study of the influence of the hyperparameters of the model, including $n$, in the following section. We must remark that other parametrizations are possible and in principle can be tuned to improve the results of each specific problem. 

\begin{figure}
    \centering
    \begin{subfigure}{\columnwidth}
        \centering
        \includegraphics[width=\columnwidth]{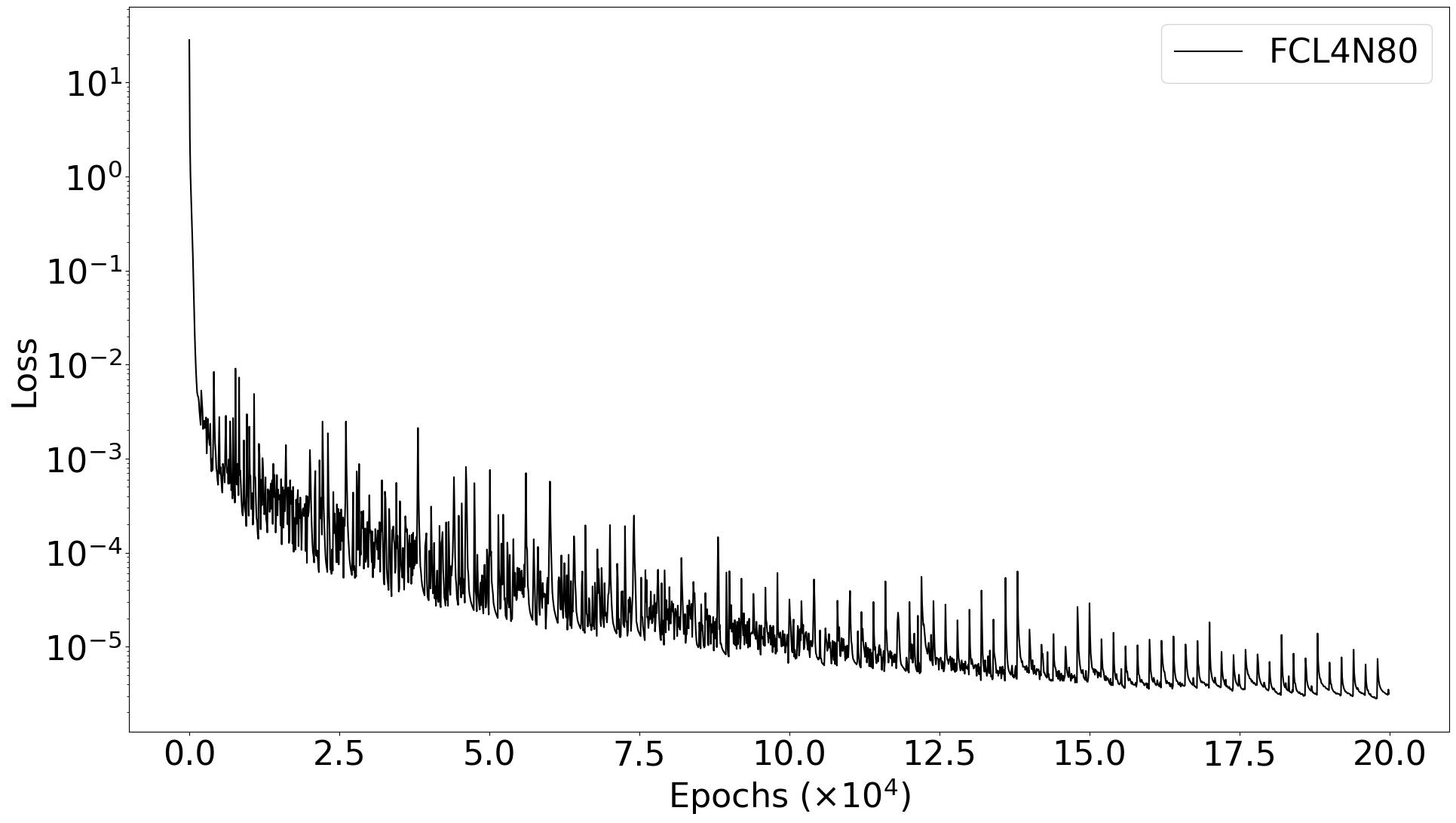}
        \caption{}
        \label{fig:FCL4N80_loss}
    \end{subfigure}
    \begin{subfigure}{\columnwidth}
        \centering
        \includegraphics[width=\columnwidth]{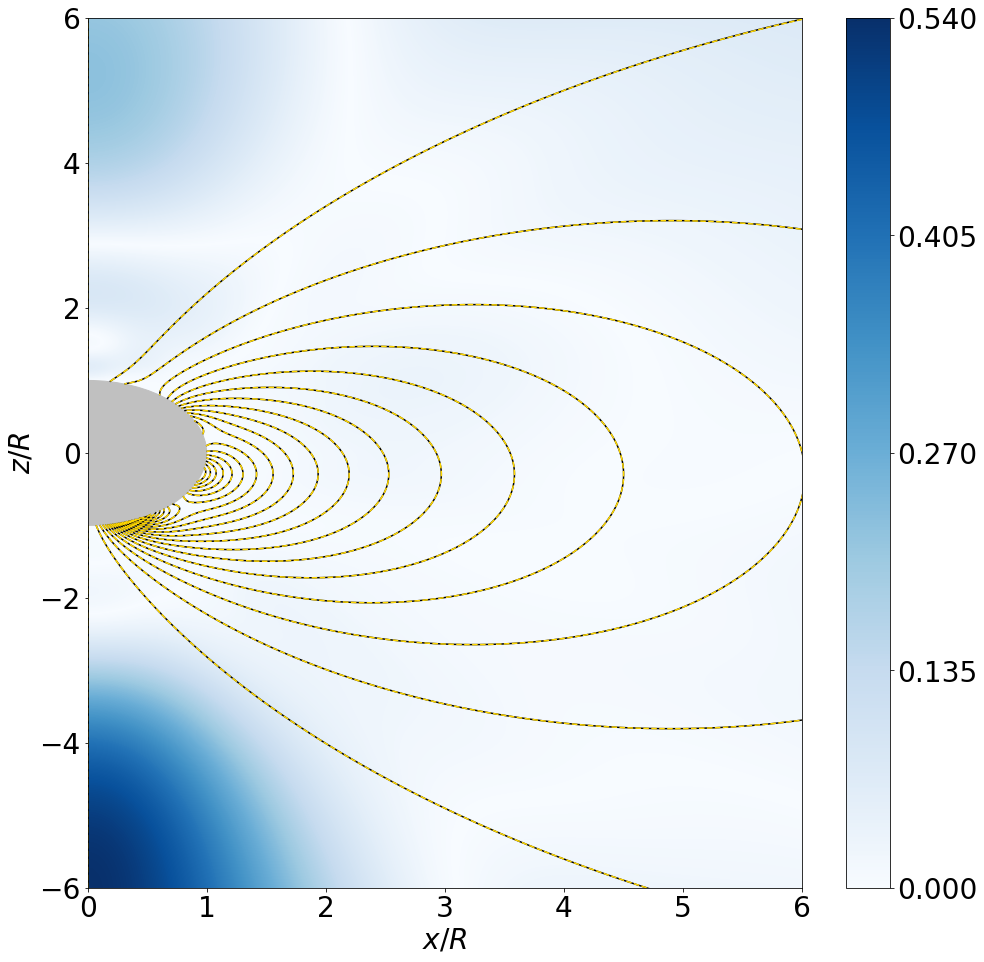}
        \caption{}
        \label{fig:FCL4N80_relative_error}
    \end{subfigure}
    \caption{(a) Evolution of the loss function with the training epochs. The periodic spikes correspond to renewals of the training set. (b) Colormap of the relative error (percentage) between $\mathcal{P}$ and $\mathcal{P}_{\text{ex}}$. Yellow dashed lines correspond to contours of $\mathcal{P}$ while black solid lines correspond to $\mathcal{P}_\text{ex}$. The multipole coefficients in $f_b$ for this particular example are $b_1 = 1$, $b_{l\geq 2} = \left(-1\right)^{l+1}0.6$.}
\end{figure}

The input layer of the NN consists of the coordinates of the point where the solution will be evaluated $(q, \mu)$ and the coefficients $b_l$ determining the BC at $q=1$.
For the physical applications in this paper, we expect the dipole ($l=1$) component of the magnetic field to be dominant. 
Therefore, we normalise all other multipole coefficients by dividing them by $b_1$ and we reduce the range of the possible values of $b_{l > 1}$ between -1 and 1. 
With this choice, we can omit the $b_1$ coefficient because it is reabsorbed in the normalisation factor of the magnetic field strength.

Hereafter we limit ourselves to $l_{\text{max}} = 7$, which suffices for our purposes and is a good compromise between generality of solutions and ease of training. Increasing the number of multipoles adds complexity and it would require larger networks to achieve the desired accuracy.
Thus, one training point is defined as
$$(q, \mu, b_{2}, b_{3}, b_{4}, b_{5}, b_{6}, b_{7}),$$

In each forward pass providing $\mathcal{N} \left(q,\mu, \left \lbrace b_{l} \right \rbrace \right)$, the solution of the differential equation at each training point $i$ is given by
\begin{equation}\label{eq:pinn_poloidal_function}
    \mathcal{P}_i = \left(1-\mu^2\right) \left[\sum_{l=1}^{l_{\text{max}}} \frac{b_{l}}{l} q^{n} P_{l}'(\mu) + q (1-q) \mathcal{N} \left(q,\mu, \left \lbrace b_{l} \right \rbrace \right) \right]. 
\end{equation}
Notice that the output of the network $\mathcal{N}$ depends both on the coordinates and on the BCs. This is the crux of our approach, as we want the network to be able to generalise for any BC (expressed in terms of the $b_l$ coefficients).

The weights and biases of the network are optimised by minimising the loss function averaged over a large sample of training points $i_{\text{max}}$
\begin{equation}\label{eq:loss_function_vacuum}
    \mathcal {J} = \frac{1}{i_{\text{max}}}\sum_{i=1}^{i_{\text{max}}}{ \left[\triangle_{\text{GS}}\mathcal{P}_i \right]^2},
\end{equation}
using the ADAM optimiser \citep{Kingma_Ba_2014} with an exponential learning rate decay. The derivatives in the Grad--Shafranov operator $\triangle_{\text{GS}}$ are calculated with the automatic differentiation tools already implemented in the machine learning framework that we use (TensorFlow \citep{tensorflow_2016}).
We consider training sets of size $i_\text{max} = 10^4$. At first sight, this number might look large. However, we must stress that it is required by the high dimensionality of our problem. Our parameter space consists of two coordinates plus six coefficients to describe the BC (fixing $b_1=1$). 
Covering this 8-dimensional parameter space with only 3 points in each dimension would need $3^8=6561$ points, 
which makes evident the crucial difference between training to solve a PDE with fixed BCs or training with arbitrary BCs (formally, an infinite number of additional parameters). The training set is changed periodically every few thousand epochs in order to feed the network with as many points as possible.

Fig.~\ref{fig:FCL4N80_loss} shows the evolution of the loss function~\eqref{eq:loss_function_vacuum} with the number of training epochs. For this particular model, we have chosen a FCNN architecture with $4$ hidden layers and $80$ neurons at each layer with $n=5$ in $f_b$ (see the next section for details on the choice of these hyperparameters).
The periodic spikes correspond to renewals of the training set. 
They can be understood as a measure of the ability of a model to generalise to new, unseen points.
Fig.~\ref{fig:FCL4N80_relative_error} shows an example of the final result once the NN has been trained. 
The black solid lines show the {\it exact} analytical solution which is uniquely determined by the $b_l$ coefficients
\begin{equation}\label{eq:vacuum_solution}
    \mathcal{P}_{\text{ex}}(q,\mu) = \left(1-\mu^2\right) \sum_{l=1}^{l_{\text{max}}} \frac{b_l}{l} q^l P_l'(\mu),
\end{equation}
while the yellow dashed lines show the solution acquired by the PINN ($\mathcal{P}$). 
They are indistinguishable at the figure scale. 
The colourmap indicates the relative difference between $\mathcal{P}$ and $\mathcal{P}_{\text{ex}}$, at most $\sim$ 0.5\% for this particular model.

The magnetic field components can be computed from $\mathcal{P}$ via automatic differentiation.
From Eq.~\eqref{eq:poloidal_toroidal_decomposiosion} we have
\begin{align}
   B_r &= -q^2\frac{\partial \mathcal{P}}{\partial \mu} , \label{eq:B_r_from_P}\\
   B_{\theta} &= \frac{q^3}{\sqrt{1-\mu^2}}\frac{\partial \mathcal{P}}{\partial q}\label{eq:B_theta_from_P}.
\end{align}

In finite difference schemes, one usually losses accuracy when taking numerical derivatives. To explore the performance of the PINN in this respect, we have computed different relative error norms of different orders ($p$) for $\mathcal{P}$, $B_r$, $B_\theta$ and for the magnetic field modulus $B = \sqrt{B_r^2 + B_\theta^2}$. Results are summarised in Tab.~\ref{tab:errors}. These $p$-norms for a given order $p$ are calculated for every variable as in
\citealt{Eivazi_Tahani_Schlatter_Vinuesa_2022} 
\begin{equation}
    E_u = \frac{\lVert u - u_{\text{ex}} \rVert_{p}}{\lVert u_{\text{ex}} \rVert_{p}} \times 100, \label{eq:p_error_norm}
\end{equation}
where $u$ is the result of each variable returned by the PINN and $u_{\text{ex}}$ is the corresponding exact solution. 
Interestingly, the errors are of the same order of magnitude for the function $\mathcal{P}$ and its derivatives.
This can be attributed to the fact that we train the PINN with a second-order PDE (the loss function involves second-order derivatives) and this includes additional information on the derivatives. Furthermore, using automatic differentiation is also an advantage over finite difference schemes, where accuracy depends on the resolution. 

\begin{table}
    \centering
    \begin{tabular}{|c|c|c|c|c|c|}
        \hline
         & $E_\mathcal{P}$ & $E_{B_r}$ & $E_{B_\theta}$ &  $E_B$ \\ \hline
        $L_1$ norm & $0.017$ & $0.017$ & $0.025$ & $0.015$ \\ \hline
        $L_2$ norm & $0.019$ & $0.023$ & $0.045$ & $0.023$ \\ \hline
    \end{tabular}
    \caption{Relative error norms (percentage) between PINN and the exact solution. The numbers shown are averages of the respective norms of $100$ new sets, consisting of $10^4$ random points each. For each set, we compute the norms using Eq.~\eqref{eq:p_error_norm}.}
    \label{tab:errors}
\end{table}

\subsection{Influence of the PINN hyperparameters}\label{sec:influence_of_the_PINN_hyperparameters}

We have performed a detailed study to measure the influence of various hyperparameters of our model. 
In particular, we have considered the following:
\begin{itemize}
    \item Changes of the parametrization of the boundary ($q^n$ power).
    \item Number of neurons at each layer. 
    \item Number of hidden layers. 
    \item Resnet vs. FC architectures.
\end{itemize}
We changed one hyperparameter at a time while keeping the rest fixed to the reference values of the previous section. 
The results of this study are presented separately in the following subsections.

\subsubsection{Changes of the parametrization of the boundary conditions}

\begin{figure}
    \centering
    \includegraphics[width=\columnwidth]{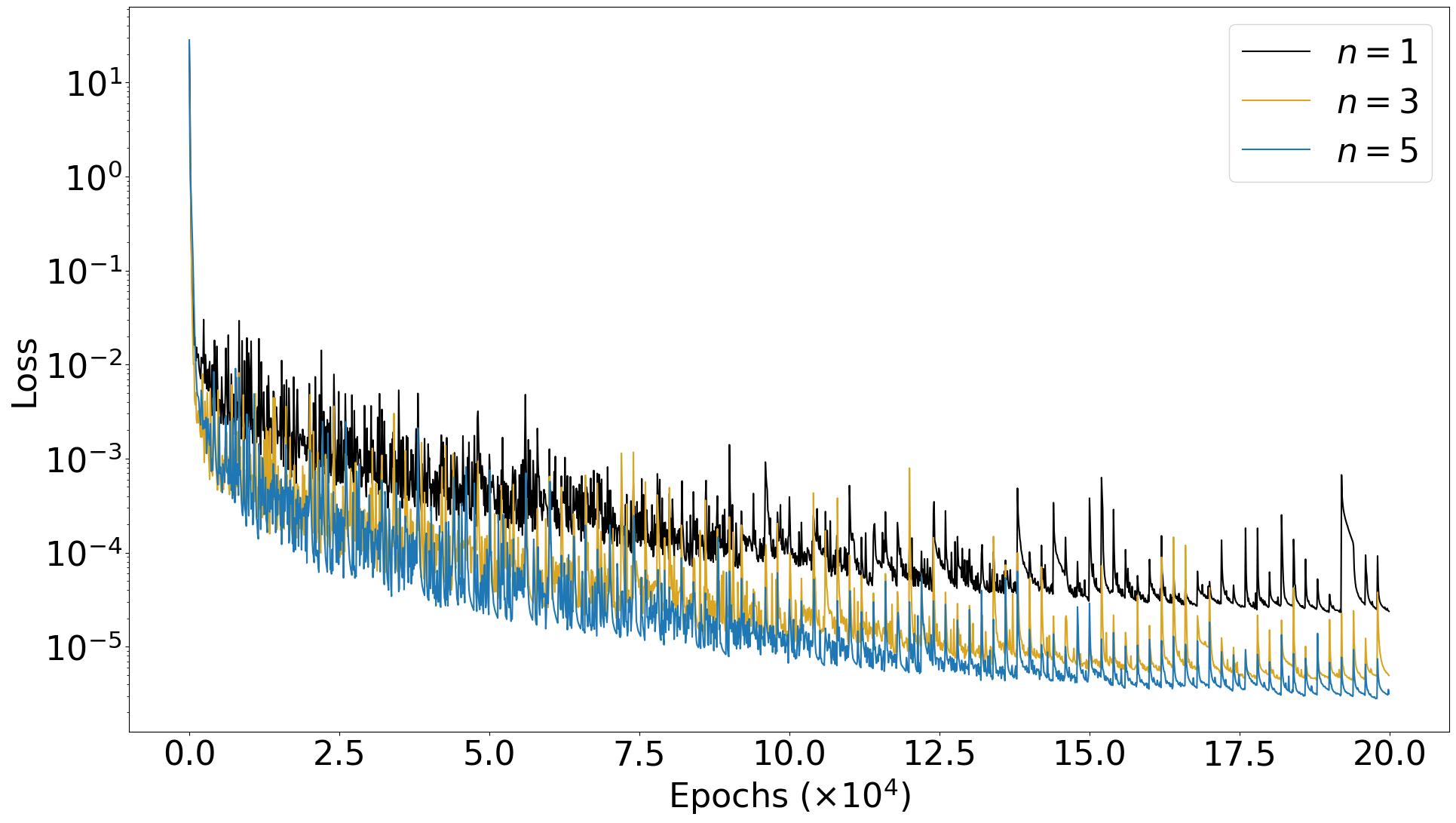}
    \caption{Evolution of the loss function with the training epochs for different values of the exponent $n$ in $f_b$. The rest of the hyperparameters are as in Sec.~\ref{subsec:current-free}.}
    \label{fig:loss_parametrization}
\end{figure}

We begin by considering different values of the exponent in the $q^{n}$ term in the boundary function~\eqref{eq:f_b}.
Fig.~\ref{fig:loss_parametrization} shows the evolution of the loss with the training epochs for three different values of the exponent $n$, namely 1 (corresponding to the Lagaris parametrization), 3 and 5. 
As $n$ increases, the impact of the surface BC becomes less important. 
In general, increasing $n$ improves the convergence of the model and leads to more accurate solutions. Tab.~\ref{tab:error_norms_parametrization} shows the relative error $L_2$-norms for the four quantities that we use to evaluate our results ($\mathcal{P}, B_r, B_\theta, B$). 
All of them decrease with increasing $n$.

\begin{table}
    \centering
    \begin{tabular}{|c|c|c|c|c|}
        \hline
        \hline
        \multicolumn{5}{|c|}{$p=2$} \\ \hline
        n & $E_{\mathcal{P}}$ & $E_{B_r}$ & $E_{B_\theta}$ & $E_B$ \\ \hline
        $1$ & $0.057$ & $0.100$ & $0.131$ & $0.084$ \\ \hline
        $3$ & $0.030$ & $0.071$ & $0.117$ & $0.064$ \\ \hline
        $5$ & $0.019$ & $0.023$ & $0.045$ & $0.023$ \\ \hline
    \end{tabular}
    \caption{
    Relative error $L_2$-norms for different values of $n$. The results indicate a higher accuracy of the overall solution as $n$ increases.}
    \label{tab:error_norms_parametrization}
\end{table}

\subsubsection{Number of neurons per hidden layer}

\begin{figure}
    \centering
    \includegraphics[width=\columnwidth]{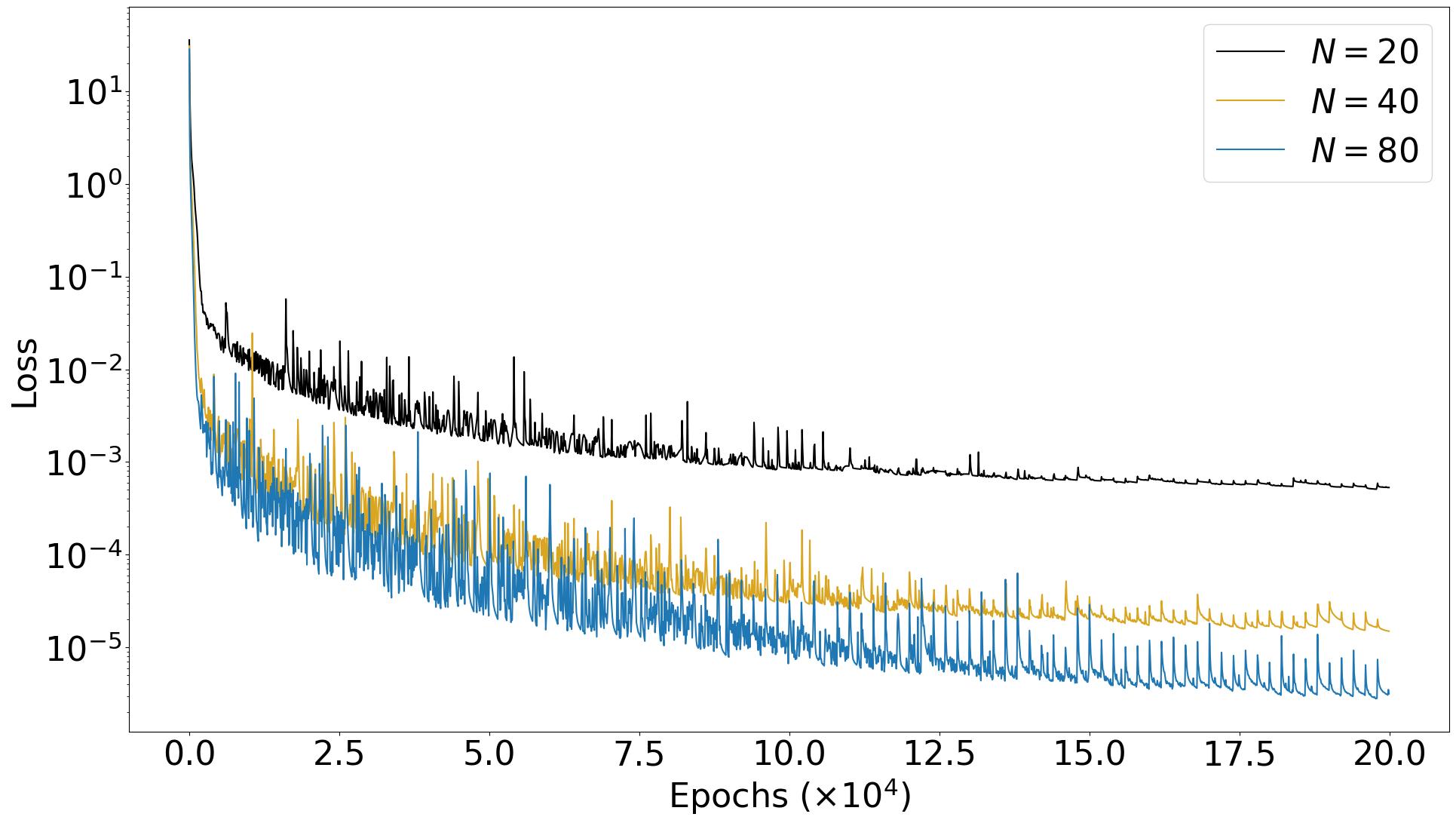}
    \caption{Evolution of the loss function with the training epochs for different values of the number of neurons per hidden layer $N$. The rest of the hyperparameters are as in Sec.~\ref{subsec:current-free}}.
    \label{fig:loss_neurons}
\end{figure}

Next, we explore the effect of the number of neurons per hidden layer $N$. 
Fig.~\ref{fig:loss_neurons} shows the evolution of the loss with the training epochs for $N = 20, 40, 80$. 
The number of neurons has a considerable impact on the convergence of each model. 
This is expected, because models with smaller $N$ do not have enough free parameters to account for the complexity and variability of the solutions. 
Our results show that the loss reaches values that are two orders of magnitude smaller when doubling $N$ from $20$ to $40$ and another order of magnitude when doubling from $40$ to $80$. 
This is reflected, as well, in Tab.~\ref{tab:error_norms_neurons}, where all quantities show a significant improvement in accuracy as $N$ increases.

\begin{table}
    \centering
    \begin{tabular}{|c|c|c|c|c|}
        \hline
        \hline
        \multicolumn{5}{|c|}{$p=2$} \\ \hline
        $N$ & $E_{\mathcal{P}}$ & $E_{B_r}$ & $E_{B_\theta}$ & $E_B$ \\ \hline
        $20$ & $0.316$ & $0.188$ & $0.285$ & $0.171$ \\ \hline
        $40$ & $0.039$ & $0.033$ & $0.052$ & $0.031$ \\ \hline
        $80$ & $0.019$ & $0.023$ & $0.045$ & $0.023$ \\ \hline
    \end{tabular}
    \caption{Relative error $L_2$-norms for different values of $N$. The results indicate a higher accuracy of the solution as $N$ increases.}
    \label{tab:error_norms_neurons}
\end{table}

\subsubsection{Number of hidden layers}

\begin{figure}
    \centering
    \includegraphics[width=\columnwidth]{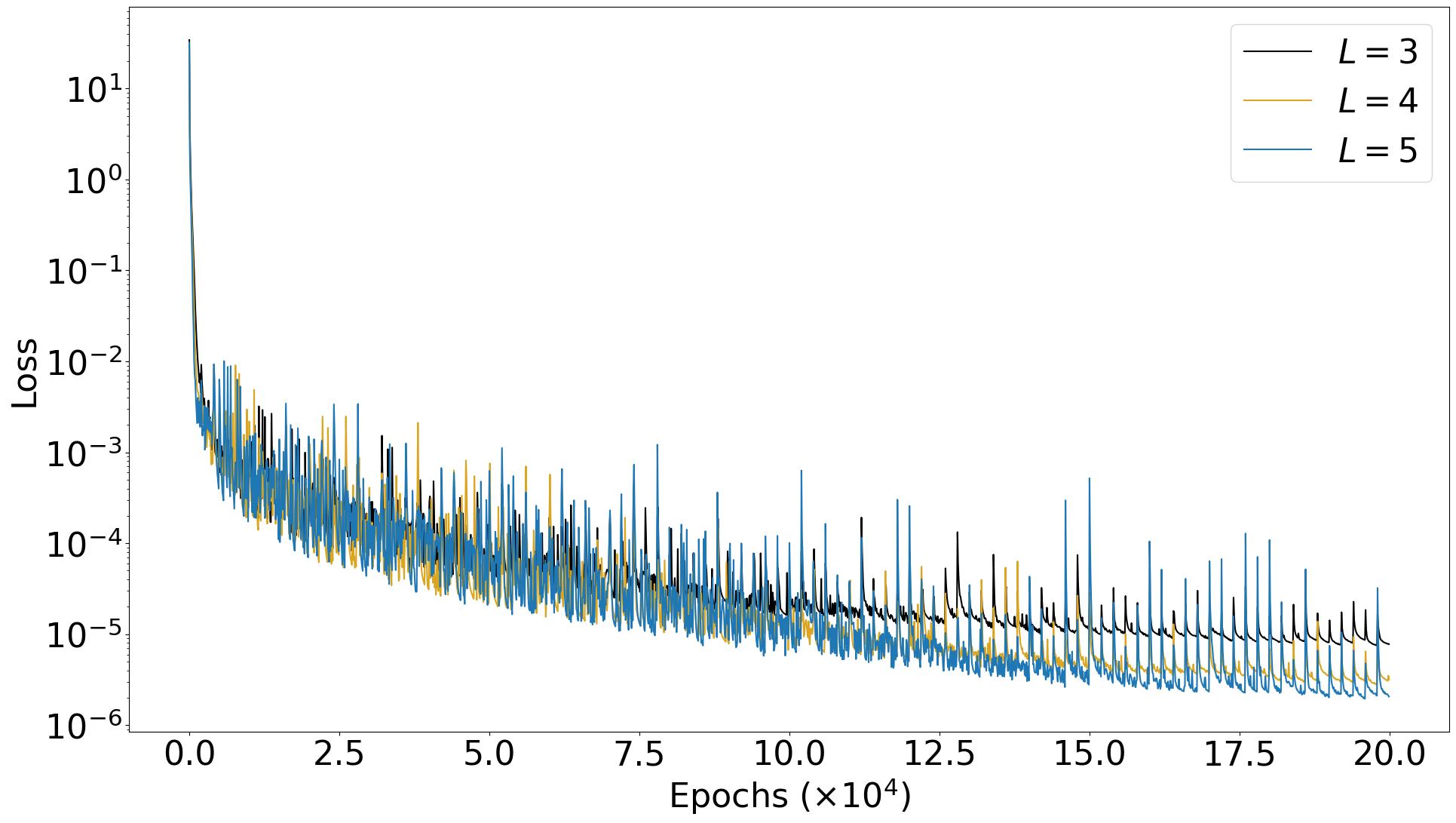}
    \caption{Evolution of the loss function with the training epochs for different values of the number of hidden layers $L$. The rest of the hyperparameters are as in Sec.~\ref{subsec:current-free}.}
    \label{fig:loss_layers}
\end{figure}

Following the same line of arguments, one could expect that increasing the number of hidden layers $L$ would also lead to improved convergence and higher accuracy. 
However, our results show that adding more layers has a marginal impact on convergence and accuracy, or it can even lead to worse results for large networks.
In other words, deeper networks are more prone to overfitting. 
Evidence of overfitting can be seen in Fig.~\ref{fig:loss_layers} for $L=5$. 
The spikes that correspond to renewals of the set of training points are much higher than expected, even at the later stages of training. 
This is, indeed, reflected in Tab.~\ref{tab:error_norms_layers}, where the model with $L=5$ performs worse in terms of accuracy than the model with $L=4$ because it is overfitted to the training set and fails to generalise to unseen points. 
Considering, in addition, that training deeper networks is computationally expensive, we conclude that increasing too much the number of layers is not beneficial in terms of accuracy or convergence.

\begin{table}
    \centering
    \begin{tabular}{|c|c|c|c|c|}
        \hline
        \hline
        \multicolumn{5}{|c|}{$p=2$} \\ \hline
        $L$ & $E_{\mathcal{P}}$ & $E_{B_r}$ & $E_{B_\theta}$ & $E_B$ \\ \hline
        $3$ & $0.030$ & $0.029$ & $0.050$ & $0.028$ \\ \hline
        $4$ & $0.019$ & $0.023$ & $0.045$ & $0.023$ \\ \hline
        $5$ & $0.014$ & $0.025$ & $0.064$ & $0.027$ \\ \hline
    \end{tabular}
    \caption{Relative error $L_2$-norms for different values of $L$. The results indicate that adding more layers does not improve the accuracy significantly and can lead to overfitting.}
    \label{tab:error_norms_layers}
\end{table}

\subsubsection{Resnet vs Fully Connected}

\begin{figure}
    \centering
    \includegraphics[width=\columnwidth]{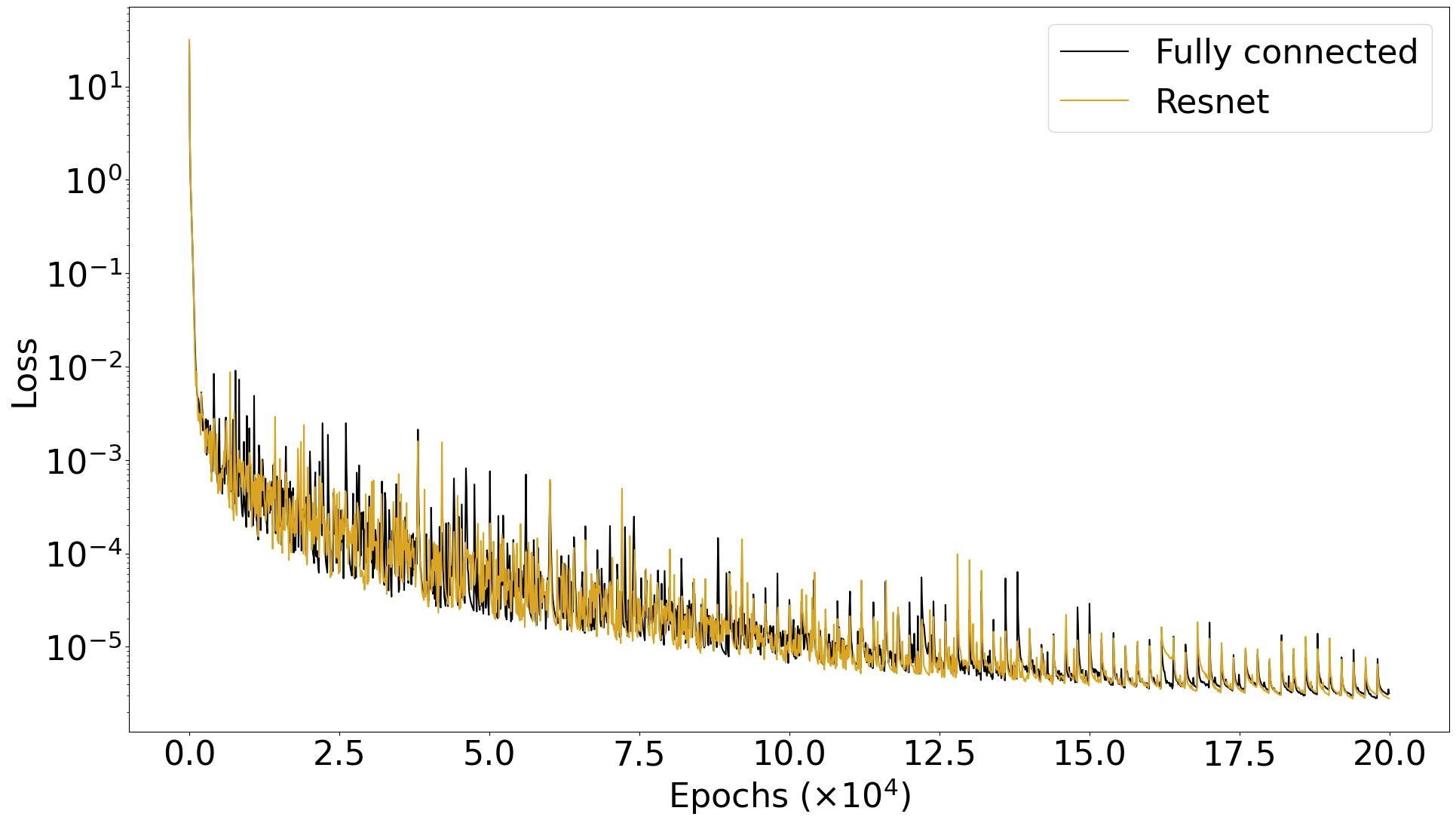}
    \caption{Evolution of the loss function with the training epochs for different NN architectures. The rest of the hyperparameters are as in Sec.~\ref{subsec:current-free}.}
    \label{fig:loss_architecture}
\end{figure}

Lastly, we consider two different types of NN architectures, FC architecture and ResNet architecture. Results are summarised in Fig.~\ref{fig:loss_architecture} and Tab.~\ref{tab:error_norms_architecture}. No appreciable differences can be detected between the two models in convergence or overall accuracy.

\begin{table}
    \centering
    \begin{tabular}{|c|c|c|c|c|}
        \hline
        \hline
        \multicolumn{5}{|c|}{$p=2$} \\ \hline
        Model & $E_{\mathcal{P}}$ & $E_{B_r}$ & $E_{B_\theta}$ & $E_B$ \\ \hline
        FCL4N80 & $0.019$ & $0.023$ & $0.045$ & $0.023$ \\ \hline
        ResL4N80 & $0.018$ & $0.019$ & $0.034$ & $0.018$ \\ \hline
    \end{tabular}
    \caption{Relative error $L_2$-norms for different NN architectures The results indicate that there is no appreciable difference for the two architectures considered.}
    \label{tab:error_norms_architecture}
\end{table}

\subsection{Force-free Grad--Shafranov equation}
\label{sec: FF PINN}
Once we have assessed the performance of our approach in the vacuum case by comparing our results with the analytical solutions, we now turn to the general case ($G(\mathcal{P})\neq 0$). 

\begin{figure}
    \centering
    \includegraphics[width=\columnwidth]{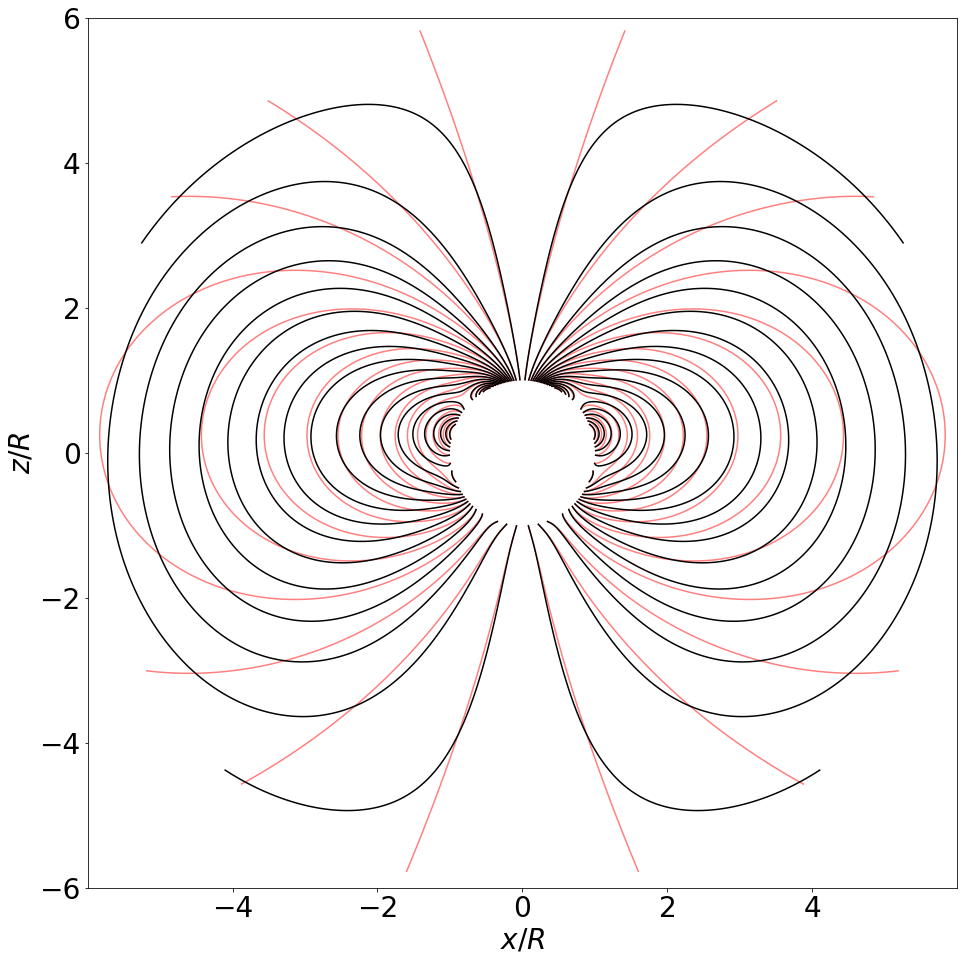}
     \caption{Field lines for the current-free (red) and force-free (black) cases. The multipole coefficients at the surface are $b_{l>1} = 0.5$ for both. For the force free case, the coefficients in expression \eqref{eq:T_of_P_quadratic} for $\mathcal{T}(\mathcal{P})$ are $s_1=0.2$, $s=0.4$.}\label{fig:current_free_vs_force_free}
\end{figure}

The configuration of the PINN solver for this case is similar to the one described in the current-free case. 
$f_b, h_b, n, l_{\text{max}}, i_{\text{max}}, N, L$, architecture, number of epochs and optimiser are the same as in Sec.~\ref{subsec:current-free}. 
There are two main differences: a) the loss function and b) the input. 
The loss function now includes the non-zero source term $G(\mathcal{P})$ which accounts for the presence of currents and is given by
\begin{equation}\label{eq:loss_function_FF}
    \mathcal {J} = \frac{1}{i_\text{max}}\sum_{i=1}^{i_{\text{max}}}{\left[\triangle_{\text{GS}}\mathcal{P}_i - G(\mathcal{P}_i) \right]^2}.
\end{equation}
The input must include information about the functional form of $G(\mathcal{P})$ or equivalently $\mathcal{T}(\mathcal{P})$. In Sec.~\ref{sec:basic_equations} we modelled $\mathcal{T}$ to be a quadratic function of $\mathcal{P}$ using two parameters, $s_1$ and $s_2$ (see Eq.~\eqref{eq:T_of_P_quadratic}).
Therefore, the input of the PINN must be extended to include these parameters. The PINN is trained to provide solutions for any value of $s_1$ and $s_2$ in the same sense that it is trained to provide solutions for any value of the multipole coefficients defining the BC. 
The input for the general force-free case is 
$$(q, \mu, b_{2}, b_{3}, b_{4}, b_{5}, b_{6}, b_{7}, s_1, s_2).$$
We note that there can be regions in the input parameter space where mathematical solutions do not exist (see \cite{Akgun_Cerda_Miralles_Pons_2018,Mahlmann_Akgun_Pons_Aloy_Cerda_2019} for a detailed discussion). 
Nevertheless, the PINN will return approximate solutions. It is up to the user to carefully evaluate the validity and accuracy of the results. Fig. \ref{fig:current_free_vs_force_free} shows, for reference, a comparison between a current-free and a force-free magnetic field, where we observe notorious difference in the structure of the field lines.

The lack of analytical solutions in the general case makes it difficult to estimate errors, which is a fundamental part of any scientific analysis. 
Despite the abundant literature on PINNs as PDE solvers in recent years, a systematic and consistent way for measuring errors and deciding on the quality of the provided approximate solutions is still lacking. Here, 
we adopt the following approach:
\begin{itemize}
    \item 
    After our NN is trained, we create a regular grid $(q_i, \mu_j)$, where $i,j = 0, ..., N_0$, and evaluate the PINN solution $\mathcal{P}$ at all the points with a forward pass. 
    \item
    Then, we discretise Eq.~\eqref{eq:grad_shafranov} using a second-order finite difference scheme, which gives us a different evaluation of the residual $\epsilon_{\text{FD}}$ using the same function values evaluated in the previous step.
    \item
    If $\mathcal{P}$ was the "exact" solution of the PDE, the second-order residual $\epsilon_{\text{FD}}$ would decrease with increasing resolution as $\sim N_0^{-2}$, where $N_0$ is the number of grid points (assuming both dimensions have the same resolution). In reality, $\mathcal{P}$ is only an approximate solution with an intrinsic error $\epsilon_{\text{NN}}$ inherited from the quality and accuracy of the PINN. Therefore, $\epsilon_{\text{FD}}$ will follow this power law only up to the point where the PINN approximation error $\epsilon_{\text{NN}}$ starts to dominate the discretisation error $\epsilon_{\text{FD}}$.
\end{itemize}

\begin{figure}
    \centering
    \includegraphics[width=\columnwidth]{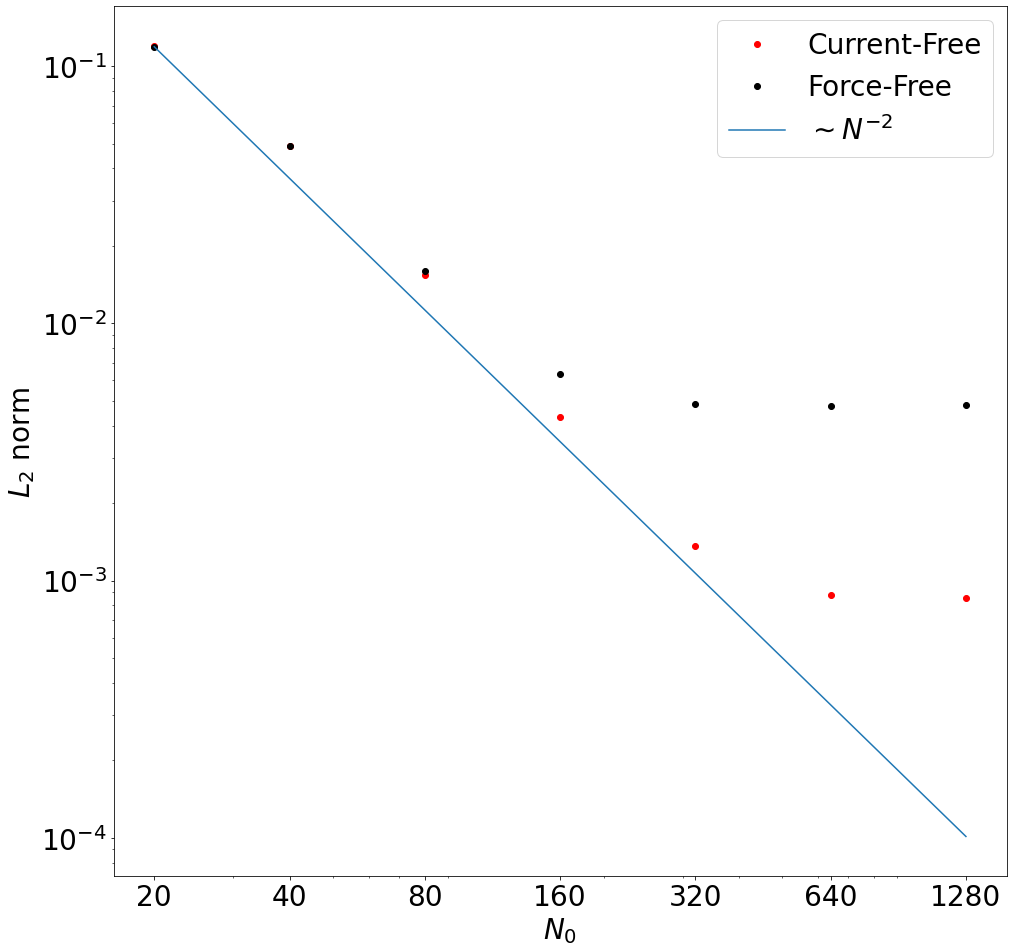} 
    \caption{The $L_{2}$-norm of the discretised Grad--Shafranov equation for the current-free (red) and force-free (black) cases as a function of the resolution $N_0$. }\label{fig:finite_diff_error}
\end{figure}

Fig.~\ref{fig:finite_diff_error} illustrates this behaviour. We plot the $L_2$-norm of the discretised GS equation
for both the current-free ($G({\cal P})=0$) and force-free cases ($G({\cal P}) \ne 0$) as a function of the number of grid points $N_0$. 
In both cases the $L_2$-norm drops as $\sim N_0^{-2}$ until it reaches a plateau which signalises that $\epsilon_{\text{NN}} > \epsilon_{\text{FD}}$. We expect that, at worst, $\epsilon_{\text{NN}}$ will be of the order of the square root of the loss function, because in Eqs.~\eqref{eq:loss_function_vacuum},~\eqref{eq:loss_function_FF} $\cal{J}$ is precisely $L_2^2$. In other words, when calculating $\epsilon_{\text{NN}}$ for a particular example, with fixed multipole coefficients and source terms, we expect the error to be of the order $\sqrt{\cal{J}}$, within a factor of a few. We note that we obtain errors of the same order of magnitude for both cases, with a factor $\sim 5$ less for the vacuum. We expect a slightly higher error when we introduce the current term $G(\mathcal{P})$, because we increase the dimensionality of the problem, and we also introduce non-linear terms into the differential equation.

\section{Application to the magnetothermal evolution of neutron stars}
\label{sec:application_to_MT}


\begin{figure*}
    \centering
    \includegraphics[width=.47\textwidth]{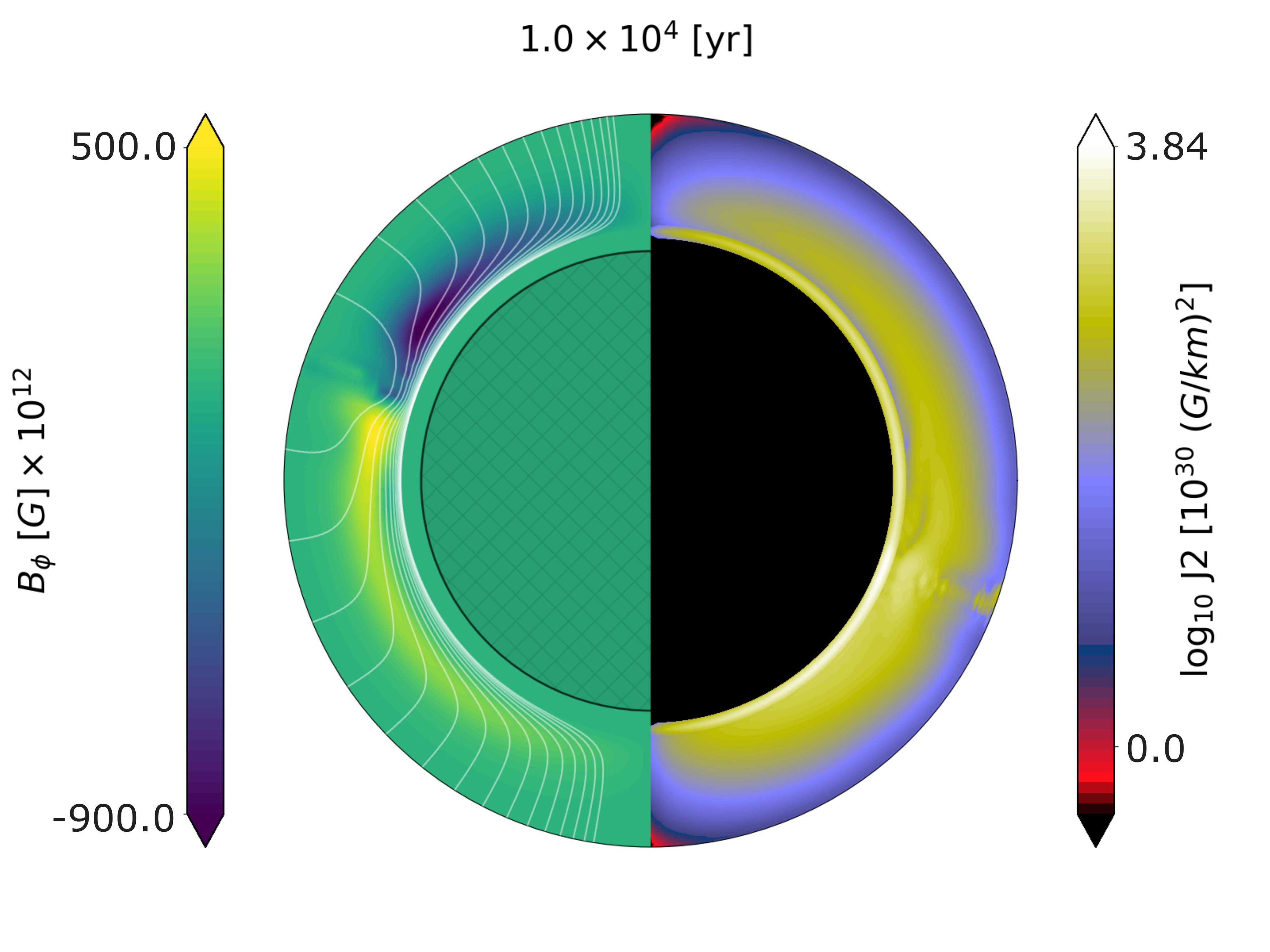}
    \includegraphics[width=.47\textwidth]{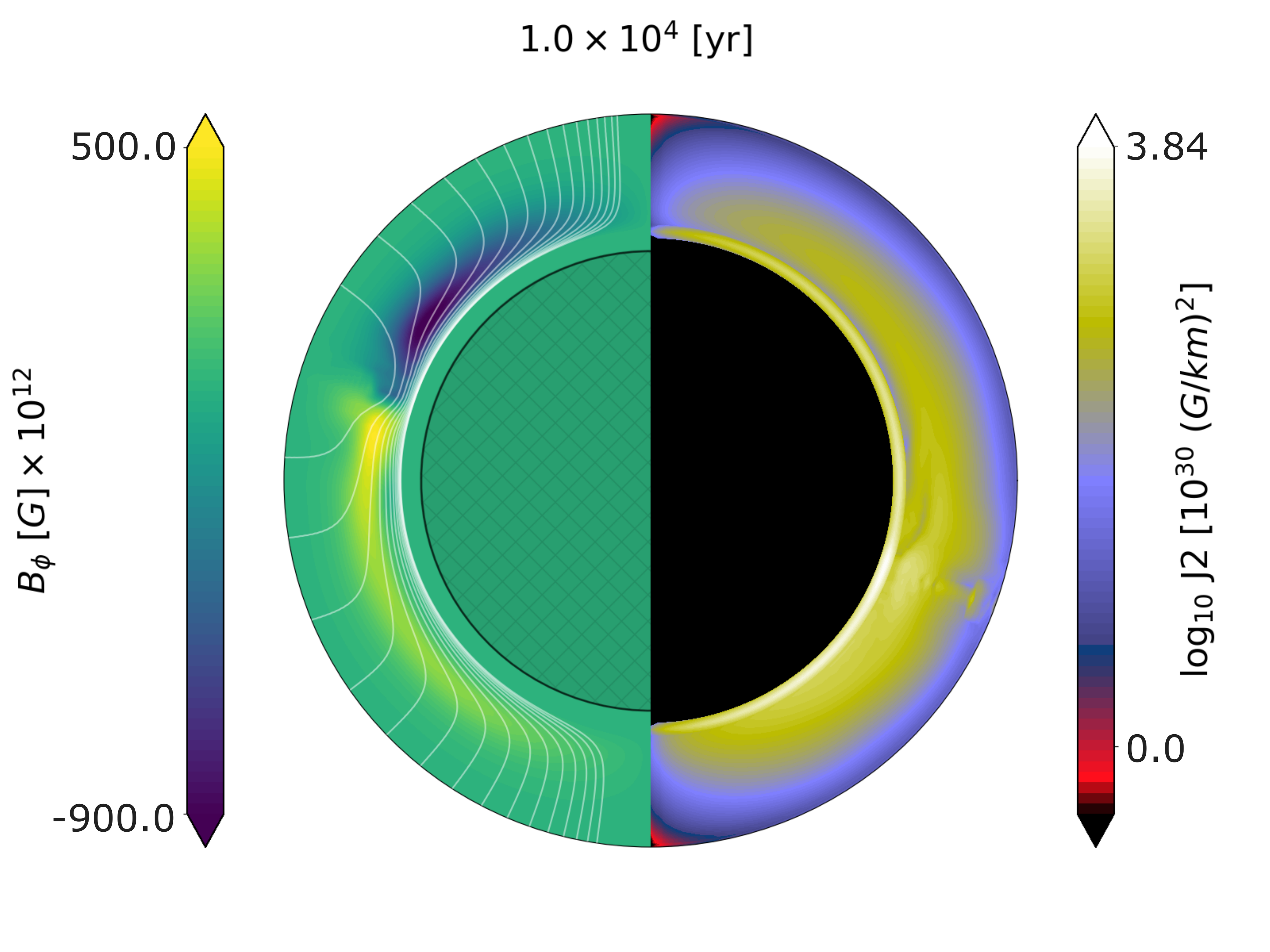}
    \caption{A snapshot of the magnetic field evolution and the electric current at $10$~kyr, obtained using OLD (left panel) and PINN (right panel). In the left hemisphere, we show the meridional projection of the magnetic field lines (white lines) and the toroidal field (colors). In the right hemishphere, we display the square of the modulus of the electric current, i.e., $|J|^2$ (note the $\log$ scale). The crust has been enlarged by a factor of 8 for visualization purposes.}
    \label{fig: B field OLD PINN}
\end{figure*}


Our astrophysical scenario of interest is the long-term evolution of magnetic fields in NSs. The evolution of the system is governed by two coupled equations: the heat diffusion equation and the induction equation (see the review by \cite{pons2019} for more details). They must be complemented with a detailed specification of the local microphysics (neutrino emissivity, heat capacity, thermal and electrical conductivity) and the structure of the star, usually assumed as fixed throughout the NS’s life. Our NS background model is a $1.4 M_\odot$ NS built with the Sly4\footnote{\url{https://compose.obspm.fr/}} equation of state \citep{douchin2001}. 
We use the 2D magneto-thermal code (latest version in \cite{vigano2021}) developed by our group suitably modified to implement the external BCs using the PINN (trained as described in the previous section) to assess its performance and potential. In particular, this implementation allows us to quickly switch and compare between vacuum BCs and the barely explored force-free BCs. To our knowledge, only the work by \cite{Akgun_Cerda_Miralles_Pons_2018} has presented results from simulations that included the effect of a magnetosphere threaded by currents. They had to implement a costly elliptic solver as a BC which slowed down the code considerably. The PINN implementation should, in principle, be much easier to change, efficient, and generalisable.

\subsection{Current-free magnetospheric boundary conditions}
\label{sec: vacuum BC}

We begin by considering a crustal-confined magnetic field topology and vacuum BCs (no electrical currents circulating in the envelope and across the surface). We enforce BCs via multipole expansion of the radial magnetic field at the surface as described in \cite{pons2009,pons2019}. 


The coefficients of the multipole expansion can be computed from the radial component of the magnetic field at the surface of the star as follows:
\begin{equation}
    b_l = \frac{2l+1}{2(l+1)} \int_0^\pi B_r(R,\theta) P_l(\cos\theta) \sin\theta d\theta.\label{bl}
\end{equation}
During the evolution, we calculate at each time step the $b_l$ coefficients using Eq.~\eqref{bl}. In the classical approach, we reconstruct the values of $B_r$ and $B_\theta$ in the external ghost cells, explicitly:
\begin{align}
    B_r &= \sum_{l=1}^{l_\text{max}}{b_l\left(l+1\right)P_l\left(\cos \theta \right)\left(\frac{R}{r}\right)^{l+2}} \label{Br_vac}, \\ 
    B_\theta &= -\sin \theta \sum_{l=1}^{l_\text{max}}{b_l P_l'\left(\cos \theta \right)\left(\frac{R}{r}\right)^{l+2}}, \label{Bth_vac}
\end{align}
where $R$ is the radius of the NS. For conciseness, we refer to this procedure by the nomenclature OLD.

In the new PINNs approach, we use the $b_l$ coefficients obtained from the Legendre decomposition as inputs to the PINN. The latter returns values of the poloidal flux function ${\cal P}$, or any required component of the magnetic field by taking derivatives (Eqs.~\eqref{eq:B_r_from_P},~\eqref{eq:B_theta_from_P}). Obviously, in this case (vacuum BCs), the PINN approach does not represent any advantage because we already know how to build the analytical solution. However, we want to ensure that the results of the simulations do not show any undesirable effects before moving on to a more complex case.

To compare the different employed techniques described above, we run axisymmetric crustal-confined magnetic field simulations using the 2D magneto-thermal code \citep{vigano2021} with a grid of $99$ angular points (from pole to pole) and $200$ radial points.
The initial field has a poloidal component of $10^{14}$ G (value at the pole and consists of a sum of a dipole ($b_1=1$), a quadrupole ($b_2=0.6$) and an octupole ($b_3=0.3$).
The initial toroidal quadrupolar component has also a maximum initial value of $10^{14}$ G. The maximum number of multipoles is fixed to $l_{max}=7$ for PINN and to $l_{max} = 50$ for OLD. The objective behind the use of different $l_{max}$ is to assess the impact of truncating the multipole number when using the PINN.

The results of the comparison at $t=10$~kyr are displayed in Fig.~\ref{fig: B field OLD PINN}. On the left (right), we show the magnetic field profiles obtained with OLD (PINN) BCs. The overall evolution of the magnetic field and the electric current is very similar. Slight differences appear due to the multipolar truncation in the PINN case. We must note that, if the same maximum number of multipoles is set for both systems, we obtain almost identical results.

\subsection{Force-free magnetospheric boundary conditions}
\label{sec:FF evolution}

To couple the internal field evolution with a force-free magnetosphere PINN-solver, we must extend the vacuum case (Sec.~\ref{sec: vacuum BC}) with additional steps. In our magneto-thermal evolution code, we impose the external BCs by providing the values of the magnetic field components in two radial ghost cells for every angular cell. In the vacuum case, once the multipolar decomposition of the radial field over the surface is known, the solution in the ghost cells can be built analytically. However, in the general case, one must solve an elliptic equation in a different grid which must be extended far from the surface to properly capture to asymptotic behaviour at long distances. This process is very costly because it must be repeated tens of thousands of time steps as the interior field evolves. 
In this situation, having trained the PINN, allows us to use it as a fast tool to provide required values of the solution in the ghost cells. We proceed as follows:
First, at each evolution time step, we must know the toroidal function $\cal{T(P)}$. For simplicity, in this application we use a quadratic function. At each time step we fit the values obtained from the internal evolution one cell below the surface. The fit provides the coefficients of the quadratic interpolation $s_1$ and $s_2$ defined in Eq.~\eqref{eq:T_of_P_quadratic}. 
Next, as described in Sec.~\ref{sec: FF PINN}, $s_1$ and $s_2$ are provided as additional input parameters to the forward pass. The PINN returns the poloidal flux function $\cal{P}$ and the components of the magnetic field needed at the ghost cells of the magneto-thermal evolution code. With this information, the internal evolution can proceed to the next time step.  

\begin{figure*}
    \centering
    \includegraphics[width=.47\textwidth]{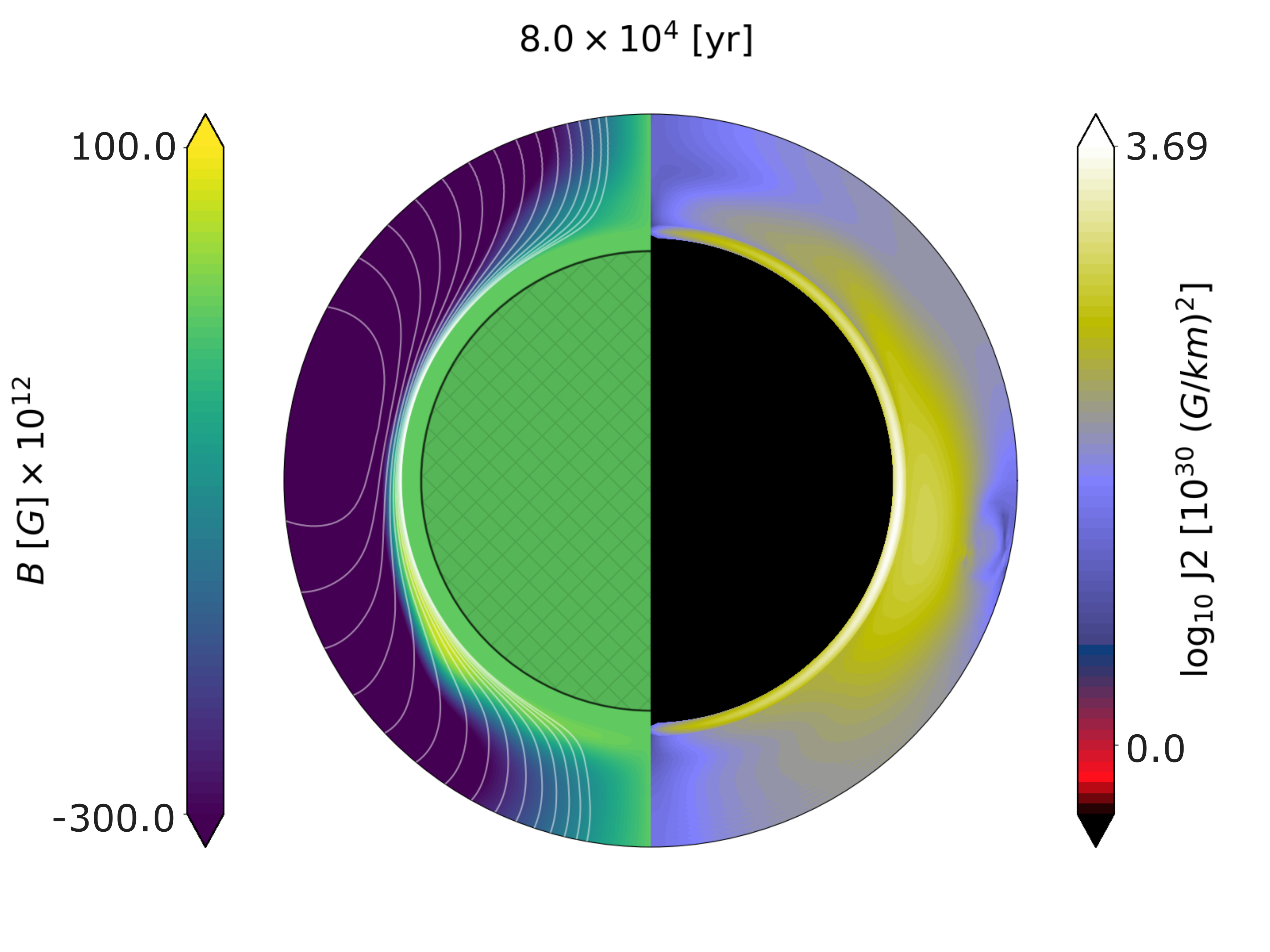}
    \includegraphics[width=.47\textwidth]{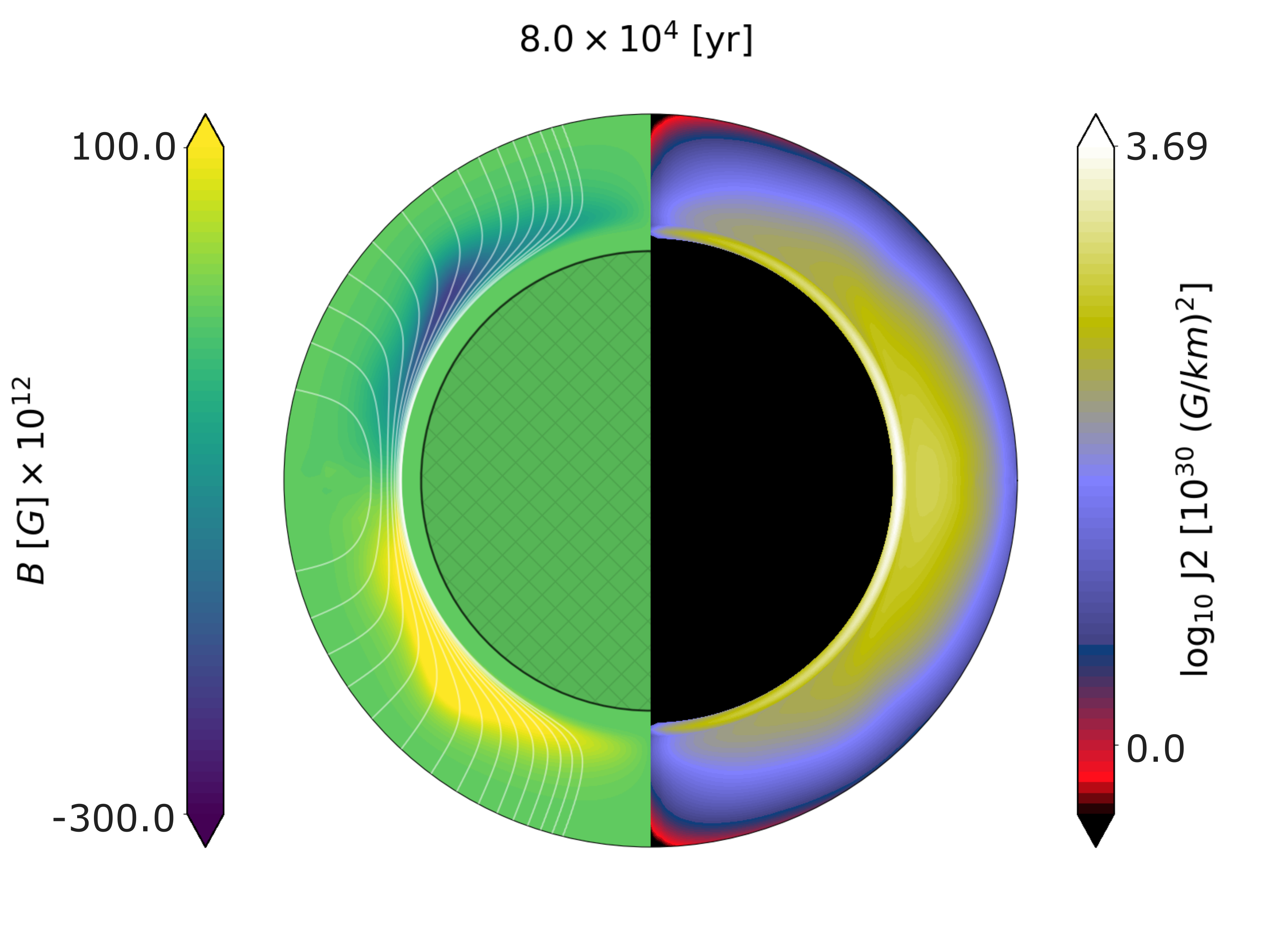}
    \caption{Same as Fig.~\ref{fig: B field OLD PINN}. 
    A snapshot of the magnetic field evolution and the electric current at $80$~kyr. Left panel: force-free BCs. Right panel: vacuum BCs.}  
    \label{fig: B field PINN FF vs VAC}
\end{figure*}

We assume an initial force-free magnetic field with a poloidal component of $3\times 10^{14}$~G at the polar surface and a maximum toroidal field of $3\times 10^{14}$~G. To understand the impact of the different BCs, we consider in one case force-free BCs (left panel of Fig.~\ref{fig: B field PINN FF vs VAC}) and in the other case vacuum BCs (right panel of Fig.~\ref{fig: B field PINN FF vs VAC}). 
The results of the comparison are illustrated by two snapshots at $t=80$~kyr of the evolution with the same initial model. 
We note that the initial force-free magnetic field allows current sheets to thread the star's surface.
A distinct magnetic field evolution is clearly observed if we apply one type of BCs or the other. The force-free BCs (left) result in a stronger toroidal dipole close to the surface and slightly displaced towards the north. The stronger toroidal component compresses the poloidal field lines closer to the poles. In contrast, for vacuum BCs,
the poloidal field lines retain certain symmetry with respect to the equator, and the dominant toroidal component is now quadrupolar and concentrated at the crust/core interface, as shown in the right panel of Fig.~\ref{fig: B field PINN FF vs VAC}. 
The distribution of the electric current in the stellar crust is also different. Enforced by the vacuum BCs, current tends to vanish around the poles and close to the surface. This is similar to what was observed in Fig.~\ref{fig: B field OLD PINN} although the initial field topology is different. 
Instead, with force-free BCs, currents near the surface are not forced to vanish. In the left panel, the slightly more yellowish region in the northern hemisphere and mid-latitudes indicates that significant current flows into the magnetosphere.  
This difference in current configurations would have important implications in the observed temperature distribution, as discussed in \cite{Akgun_Cerda_Miralles_Pons_2018}. We will address, in future works, a more detailed exploration of the effect of BCs since our purpose here is to illustrate with a few examples the potential of our approach.

\section{Conclusions}\label{sec:conclusions}

Using PINNs to obtain a solution of a particular boundary value problem is, up to date, far more computationally expensive and arguably less accurate than using classical methods. The drawbacks are related to the training process, which involves the minimization of a high-dimensional loss function. Once a PINN is trained for a given boundary value problem, its utility is limited because it would be necessary to re-train to generate new solutions with different BCs.

The functionality of PINNs to real physical problems would become significantly better if their flexibility and adaptability can be increased. 
In this work, we explore this idea by training our PINN for general BCs and source terms, expressed through appropriate coefficients (a limited number of them) that enter as additional inputs in the network. 
Of course this makes the training process computationally more expensive, but the evaluation of new generic solutions is very fast.
In our study, the coverage of the parameter space is not exhaustive because we have used very limited computational resources (a personal computer), and our purpose is to show that this proof-of-concept works and can already be applied to some physical problems, even by non-experts in computer science.
If necessary, it is straightforward to adapt our implementation for the specific needs of other applications (for example, adding more multipoles in the boundary if we need to capture smaller scales, or extending the parametrization of the BCs). We are aware that
it is possible to drastically improve the efficiency of our computations by employing GPU clusters or enriching our algorithms with advanced machine learning techniques and that will be required, for example, in the extension to 3D of this work.

We have also explored various configurations for our network through a basic hyperparameter space study.
We conclude that the most impactful element is the number of neurons per layer. 
On the contrary, making the network deeper by adding more layers is not beneficial, beyond a reasonable minimum.
Furthermore, the way that the solution is parametrized to always satisfy the BCs is important, indicating that the human orientation in some choices (as opposed to a zero-like approach) is still critical for physics problems.
We also found that ResNet architectures do not offer any advantage for the kind of applications that we are dealing with.
As is often the case in research related to NN, all the above conclusions are empirical and
should be taken with a grain of salt because it is hard to find a rigorous
theoretical foundation to support them.

Our results show that the PINN solutions are relatively accurate, reliable, and well-behaved. For the current-free Grad--Shafranov equation, where comparisons with an analytical solution can be made, we found relative differences of typically less than $1\%$.
For the force--free Grad--Shafranov equation, we propose a method for estimating the error through the use of a finite difference discretisation scheme. 
Our analysis shows that solutions are accurate up to a point that we associate to the intrinsic approximation error of the PINN.
This approach is straightforward to implement and self-consistent and could be set as the standard procedure for estimating errors of PINN-based PDE solvers in general.
Even if PINNs fall short in terms of precision compared to classical PDE solvers, they can still be used in conjunction with them in various cases. 
For example, many iterative solvers rely on good initial guesses to converge. A PINN can provide such an initial guess to be subsequently refined by a classical method. 

The most interesting case where PINNs overcome the capabilities of classical methods are physical systems with two or more domains that are governed by vastly different physical conditions and time-scales. 
An example of such a case is the magnetothermal evolution in the interior of a NS that is connected to a force-free magnetosphere. Solving this problem through a global simulation in the entire domain is very costly due to the needs of the elliptic solver for the exterior solution. 
On the contrary, PINNs provide a very effective way of imposing BCs at the interface of the two domains. 
Once a PINN is trained, accurate enough magnetospheric solutions in a few points (ghost cells of the interior evolution code) can be swiftly computed at each time step without adding an excessive amount of computational cost.
We have shown that for the well-tested case of vacuum BC, the PINN is accurate enough to satisfactorily reproduce the reference results obtained by the exact spectral decomposition approach. Nevertheless, it is about 2 times slower. As proof-of-concept, to demonstrate the potentiality of our approach, we also presented results for the less explored force-free magnetospheric BCs. In this case, the computational cost was more than an order of magnitude smaller than the similar problem solved with classical methods. 
Indeed, we obtain solutions that allow currents that thread the NS's surface and flow into the magnetosphere, giving rise to a new family of internal field configurations.
We reserve a more detailed analysis of the physical properties of these solutions in 2D, and the more physically relevant extension to 3D, for future works.

\section*{Acknowledgements}

We acknowledge the support through the grant PID2021-127495NB-I00 funded by MCIN/AEI/10.13039/501100011033 and by the European Union, and the Astrophysics and High Energy Physics programme of the Generalitat Valenciana ASFAE/2022/026 funded by MCIN and the European Union NextGenerationEU (PRTR-C17.I1). JFU is supported by the predoctoral fellowship UAFPU21-103 funded by the University of Alicante. CD is supported by the ERC Consolidator Grant “MAGNESIA” No. 817661 (P.I. N. Rea) and this work has been carried out within the framework of the doctoral program in Physics of the Universitat Aut\`onoma de Barcelona and it is partially supported by the program Unidad de Excelencia Mar\'ia de Maeztu CEX2020-001058-M.

\section*{Data Availability}

All data produced in this work will be shared on reasonable request to the corresponding author.



\bibliographystyle{mnras}
\bibliography{Bibliography} 








\bsp	
\label{lastpage}
\end{document}